\documentclass[12pt]{article}
\usepackage[pdftex]{graphicx}
\usepackage{color}
\usepackage{amsmath}
\usepackage{amssymb}
\usepackage{euscript}
\usepackage{multirow}
\usepackage{color}
\usepackage{colortbl}
\usepackage{calc}
\usepackage{cite}

\makeatletter
\newlength\twolinebox@linelength
\newlength\twolinebox@columnheight
\newcommand{\twolinebox}[2]{%
   \setlength{\twolinebox@linelength}%
             {\maxof{\widthof{#1}}{\widthof{#2}}}%
   \setlength{\twolinebox@columnheight}{\heightof{#1}+\depthof{#1}+0.2em+0.4em/2+\heightof{0}/2}%
   \raisebox{0pt}[\twolinebox@columnheight][\heightof{\vbox{\vskip0.2em\hbox to 
   \twolinebox@linelength {#1\hfil}\vskip0.4em\hbox to 
   \twolinebox@linelength {#2\hfil}}}+\depthof{\vbox{\vskip0.2em\hbox to 
   \twolinebox@linelength {#1\hfil}\vskip0.4em\hbox to 
   \twolinebox@linelength {#2\hfil}}}-\twolinebox@columnheight+0.2em]{\vbox to 
   \twolinebox@columnheight{\vskip0.2em\hbox to 
   \twolinebox@linelength {#1\hfil}\vskip0.4em\hbox to 
   \twolinebox@linelength {#2\hfil}}}%
}

\newcommand\WINHAC[0] {\textsf{WINHAC}}

\newcommand\PYTHIA[0] {\textsf{PYTHIA}}
\newcommand\PHOTOS[0] {\textsf{PHOTOS}}
\newcommand\KKMC[0] {\textsf{KKMC}}
\newcommand\LHAPDF[0]{\textsf{LHAPDF}}

\newcommand{\pTl}    {{p_{T,l}}}

\def\slashii#1{\setbox0=\hbox{$#1$}            
  \dimen0=\wd0                                 
  \setbox1=\hbox{\sl/} \dimen1=\wd1            
  \ifdim\dimen0>\dimen1                        
     \rlap{\hbox to \dimen0{\hfil\sl/\hfil}}   
     #1                                        
  \else                                        
     \rlap{\hbox to \dimen1{\hfil$#1$\hfil}}   
     \hbox{\sl/}                               
  \fi}

\definecolor{rltbrightred}{rgb}{1,0,0}
\definecolor{rltred}{rgb}{0.75,0,0}
\definecolor{rltdarkred}{rgb}{0.5,0,0}
\definecolor{rltbrightgreen}{rgb}{0,0.75,0}
\definecolor{rltgreen}{rgb}{0,0.5,0}
\definecolor{rltdarkgreen}{rgb}{0,0,0.25}
\definecolor{rltbrightblue}{rgb}{0,0,1}
\definecolor{rltblue}{rgb}{0,0,0.75}
\definecolor{rltdarkblue}{rgb}{0,0,0.5}
\definecolor{webred}{rgb}{0.5,.25,0}
\definecolor{webblue}{rgb}{0,0,0.75}
\definecolor{webgreen}{rgb}{0,0.5,0}
\definecolor{Black}{rgb}{0,0,0}
\definecolor{Greymax}{rgb}{0.65,0.65,0.65}
\definecolor{Greycen}{rgb}{0.75,0.75,0.75}
\definecolor{Greymin}{rgb}{0.85,0.85,0.85}
\definecolor{hl}{rgb}{
                0.909803922,       
                0.82745098,               
                0.909803922}

\begin{document}

\begin{titlepage}
 
\mbox{}
\vspace{5mm}

\begin{center}
{\LARGE\bf QED radiative corrections and their impact 
\vspace{2mm}\\ 
on $H \rightarrow \tau\tau$  searches at the LHC$^*$}
\end{center}

\vspace{8mm}

\begin{center}
{\large\bf  
Mieczyslaw Witold Krasny$^{a},$ 
Stanis{\l}aw Jadach$^{b}$
\vspace{2mm}\\ 
{\rm and} 
\vspace{3mm}\\ 
Wies{\l}aw P{\l}aczek$^{c}$
}

\vspace{5mm}

{\em $^a$Laboratoire de Physique Nucl\'eaire et des Hautes Energies, \\
          Universit\'e Pierre et Marie Curie -- Paris 6, Universit\'e Paris Diderot -- Paris 7, \\
          CNRS--IN2P3, 4 pl. Jussieu, 75005 Paris, France.}
\\  \vspace{2mm}
{\em $^b$Institute of Nuclear Physics, Polish Academy of Sciences,\\
  ul.\ Radzikowskiego 152, 31-342 Krak\'ow, Poland.}         
\\  \vspace{2mm}
{\em $^c$Marian Smoluchowski Institute of Physics, Jagiellonian University,\\
         ul.\ {\L}ojasiewicza 11, 30-348 Krak\'ow, Poland.}
\end{center}

\vspace{18mm}
\begin{abstract}
 \noindent  
In this paper we show  that  the excess of the $ \tau \tau $ events  
with respect to the Standard Model  background predictions, observed by the ATLAS and CMS 
collaborations  and interpreted as the evidence of the  Higgs-boson decay 
into a pair of $\tau$-leptons, may  be accounted for by properly taking into account QED radiative corrections
in the modelling of the $Z/\gamma^* \rightarrow \tau\tau$  background. 
 
\vspace{3mm}
 \end{abstract}

\vspace{7mm}
\begin{center}
{\it version published in  Eur. Phys. J. C (2016) 76:194}
\end{center}

\vspace{25mm}
\footnoterule
\noindent
{\footnotesize
$^{\star}$The work is partly supported by the program of the French--Polish 
co-operation between IN2P3 and COPIN no.\ 05-116,
and by the Polish National Centre of Science grant no.\ DEC-2012/04/M/ST2/00240.
}
\end{titlepage}

\section{Introduction}
\label{introduction}

The extraction of a $ H   \rightarrow   \tau \tau $  signal from the measured mass spectrum of $\tau\tau$ pairs 
produced at the Large Hadron Collider (LHC) requires very precise knowledge of the  expected  $ H   \rightarrow   \tau \tau $
and $Z/ \gamma^* \rightarrow \tau \tau$ mass spectra.  If the mass of the $\tau\tau$ pairs is reconstructed with perfect resolution 
then analysis is simple. However, in reality the analysis is complicated by the missing energy carried by neutrinos 
from $\tau$-decays, the resolution of the measurements of the energies of the various particles in the final state, and, last but not the least, the adequate modelling of the photon radiation processes.

In this paper we discuss the robustness of experimental evidence of the $ H   \rightarrow   \tau \tau $  decays
extracted from data collected at the LHC -- reported by the ATLAS and CMS collaborations  \cite{ATLAS_tau,CMS_tau} and 
considered as the  direct confirmation  of the Higgs-boson coupling to fermions. 
We focus our investigation on  the processes of radiation of photons by muons and $\tau$-leptons.
These processes become particularly  important  for  the Higgs-signal search  strategy in which the dominant 
$Z/ \gamma^* \rightarrow \tau \tau$ background is modelled using  $Z/ \gamma^* \rightarrow \mu \mu$ events. 
This modelling, customarily called {\it the embedding procedure}, was used by both the ATLAS and CMS 
collaborations in their respective analyses \cite{ATLAS_tau,CMS_tau}.

The embedding procedure  exposes the analyses to the  necessity of taking into account and applying the 
QED radiative corrections,  which are specific to this procedure.  The study presented 
in this paper demonstrates that they  are large and could explain all the excess of events 
interpreted as the $H   \rightarrow   \tau \tau$  signal. 

This paper is organised as follows. 
The ATLAS and CMS embedding procedure is summarised  in Section \ref{embedding}. 
Section \ref{corrections} starts with  an introductory discussion of the radiative  corrections
to the $Z/\gamma ^*$-boson leptonic decays. This discussion is  followed by a  presentation of the  Monte Carlo
tools used in the analysis of  the QED radiative processes associated with the $Z/\gamma^*$-boson 
decays into $\tau$ and $\mu$ pairs. We analyse  the dependence of radiative corrections 
on the assumed parton distributions, 
fragmentation model, handling of processes involving multi-photon emissions, in order  to  asses 
the theoretical precision of the results presented in subsequent sections of this paper.  
Finally, in Section \ref{corrections} we discuss the size of the radiative corrections for an 
analysis based on the embedding procedure and its  dependence on the experimental 
resolution of the reconstructed invariant mass of $\tau ^+ \tau ^-$ pairs. 
In Section \ref{bias} we present  our predictions  for the excess of events by properly taking into account radiative photons
in the modelling of the $Z/\gamma^* \rightarrow \tau\tau$  background.
We compare this excess with that of the predicted 
Higgs-boson signal. In our study we follow  the published ATLAS and CMS analyses  \cite{ATLAS_tau,CMS_tau}.
In Section \ref{overlooked} we discuss why the radiative corrections effects were 
not given sufficient attention in the ATLAS and CMS analyses.
In Section \ref{forward} we propose a list of 
tests that can check experimentally the size of the radiative correction effects
and discriminate between signatures of Higgs-boson  decays versus missing  radiative corrections
in the modelling of the $Z/\gamma^* \rightarrow \tau\tau$ background.
The conclusions of the paper are presented in Section~\ref{Conclusions}.

\section{Modelling of $Z/\gamma^*  \rightarrow \tau\tau$  processes with $\tau$-embedded $Z/\gamma^* \rightarrow \mu\mu$  events}
\label{embedding}

The $\tau\tau$-decays of $Z/\gamma^*$-bosons produced by the Drell--Yan process  are  the dominant source 
of background for $H \rightarrow \tau \tau$  searches at the LHC. The background evaluation method  is common 
for both ATLAS and  CMS analyses \cite{ATLAS_tau,CMS_tau}. The $Z/\gamma^*  \rightarrow \tau\tau$ decays 
are  modelled using {\it embedded event samples} in  loosely selected $Z/\gamma^*\rightarrow \mu\mu$ data. 
These event samples are  recorded for  each  data taking period. 
In the embedding procedure muon tracks and their associated energy depositions in the calorimeters  
are replaced by a  simulated detector response to  the final-state particles of 
the corresponding $\tau$-lepton decays. 

In deriving the momentum vectors of the embedded $\tau$-leptons, the momentum vectors of the corresponding muons
are used. These vectors are determined using the  muon tracks reconstructed from the central tracker and 
muon spectrometer hits. The  $\tau$-momentum  vectors  are calculated  
by taking the energies of the corresponding muons  
and correcting the $\mu$-momentum  vectors for the effect of the muon and $\tau$-lepton  
mass difference. The energy depositions in the calorimeter  which are identified as being  associated with the  
outgoing muons are removed. 
Next,  the detector response to  the $\tau$-leptons is simulated. In the initial  simulation step the 
radiation of photons and the decays of $\tau$-leptons are generated. In the final  step the response 
of the tracker and of the calorimeters  to the $\tau$-lepton decay products are simulated. 

There are several advantages of the embedding technique: the sensitivity to the Monte Carlo modelling aspects of the hadronic system associated 
with the $Z/\gamma^*$-boson production  is minimised, the pile-up effects are automatically taken into account in the 
background simulation,   including the underlying event activity. 
However, as discussed in the next section, there is a price to pay for such a simplification of the background 
estimation technique. 

 \section{Radiative corrections in $Z/\gamma ^*$  leptonic decays}
\label{corrections}

\subsection{Initial discussion}
\label{initial}

Leptons produced in  $Z/\gamma^*$-boson decays radiate photons. 
The probability of radiation depends on the lepton mass. Electrons 
radiate photons more frequently than muons, muons more frequently 
than $\tau$-leptons. The difference of probabilities is  
proportional to the logarithm of the squares of the leptons masses, 
$ (\alpha/ \pi)  \ln (m_{i}^2 / m_{j}^2 )$.

In order to describe  the observed invariant mass distributions of  pairs of leptons 
produced in $Z/\gamma ^*$-decays, the processes of final-state photon radiation (FSR)  
have to be modelled and the corresponding radiative corrections should be taken 
into account. 

The size of the radiative corrections depends not only on the lepton type, but also on the 
lepton detection apparatus, the experimental cuts  and the reconstruction method of 
the lepton momentum vector. 
Radiative corrections for the  measurements of the lepton momentum which  include 
information of radiated photons  reconstructed in the calorimeter are sizeably smaller than those 
based on its momentum vector solely measured from the curvature of the lepton track. 

In this paper we use the term {\it bare leptons} for 
those which have their momentum vector  reconstructed solely from the track parameters and  the term {\it dressed leptons} for an idealised case in which all the radiative photons are assigned to their lepton emitters 
and the photon  measured momenta are used in the  reconstruction of the  initial momentum vector 
of the leptons\footnote{Such an association, to be exact,  can only  be made 
at the amplitude,  rather than cross-section level.}.
The radiative corrections for the bare-lepton-based measurements 
are large, while the ones based on the dressed leptons are small. 
In general, the size of the radiative corrections may vary within the above boundaries. Their numerical  values 
are detector- and analysis-dependent, and reflect the efficiency  of algorithms, which 
assign the photon energy depositions in the calorimeters  to the respective outgoing leptons. 

The photon radiation by $\tau$-leptons  has an additional subtlety with respect to photon radiation by stable electrons
and long-lived muons. Photons may originate both  from the $\tau$-lepton radiation and  
from the $\tau$ decays. Since neutrinos produced in the  $\tau$ 
decays remain undetected, it is hardly possible to experimentally identify the photon origin
on an  event-by-event basis from the reconstructed  $\tau$-mass. 

At first sight the QED radiative effects   should play a negligible role in the determination 
of the strength of the $H \rightarrow \tau\tau$ decay signal. This is perhaps why 
the necessity to evaluate  the radiative correction to the 
embedding procedure escaped attention. 
As discussed in Subsection~\ref{false-signal}, these corrections,  for a
realistic detector resolution and for the adapted measurement strategy, are large 
and cannot be neglected.  

In advance of a quantitative analysis of the radiative corrections
to the embedding procedure, we first present the  Monte Carlo 
tools used in our analysis and evaluate  their precision.

 \subsection{Theoretical control} 
 \label{control}

The inclusive cross section for the $H \rightarrow \tau \tau$ decays is, in the  LHC energy range,   
three orders of magnitude lower than that for $Z/\gamma^* \rightarrow \tau \tau$  decays.
This 10$^{-3}$ ratio defines the precision {\it yard-stick} for the requisite theoretical and  
experimental control of the $Z/\gamma^* \rightarrow \tau \tau$ background. 

In the studies presented in this paper we use the Monte Carlo event generator 
\WINHAC~\cite{WINHAC:MC,Placzek2003zg,Placzek:2009jy,Placzek:2013moa}. This 
program simulates the single $W^{\pm}/Z/ \gamma^*$-boson production 
with leptonic decays  in hadronic collisions, i.e.\ the charged-current (CC) and neutral-current (NC) 
single Drell--Yan (DY) processes.  The parton-level hard processes are convoluted
with parton distribution functions (PDFs) supplied through the \LHAPDF\ library \cite{Whalley:2005nh}. 
For the $W$-boson processes \WINHAC\ features the ${\cal O}(\alpha)$ 
Yennie--Frautschi--Suura (YFS)  exclusive exponentiation for the electroweak corrections \cite{Placzek2003zg,Placzek:2009jy,Placzek:2013moa}. 
This is not the case for the $Z/\gamma^*$ processes, considered in this paper, which 
are included only at the Born level and the QED radiative corrections are provided through an interface to the 
\PHOTOS\ generator~\cite{Barberio:1994qi,Golonka:2005pn,Golonka:2006tw,Davidson:2010ew}. 
\PHOTOS\ generates multi-photon final-state radiation (FSR) in $Z/\gamma^*$ decays with the use of
an iterative algorithm in which resummation of leading QED radiative effects is improved 
with a matrix-element (ME) correction to account for sub-leading, mainly  ${\cal O}(\alpha)$, effects. 
It is  shown in Ref. \cite{Arbuzov:2012dx} that \PHOTOS\ reproduces very well the exact 
${\cal O}(\alpha)$ FSR corrections.
In this paper we refer to the generator based on the \WINHAC\  and \PHOTOS\  interface 
as \WINHAC$+$\PHOTOS.

\WINHAC\ includes only the leading-order (LO) QCD matrix elements. However,
in order to generate realistic kinematical distributions of  $Z/\gamma ^*$, it is interfaced with 
the \PYTHIA~{\sf 6.4}  generator \cite{Sjostrand:2006za}, 
which performs the initial-state QCD (and QED) parton shower, the appropriate proton-remnant treatment and
the necessary hadronisation/decays. This interface provides an improved generation of lepton transverse momenta with respect to the original  \PYTHIA~{\sf 6} program, which results in a good agreement with the NLO QCD predictions;  for details see Ref.  \cite{Krasny:2012pa}.

In the studies presented in this paper we use  as the base-line the {\sf CTEQ6.1} parametrisation
of PDFs \cite{Stump:2003yu}.
Other  parametrisations,  such  as {\sf MSTW\-2008NNLO} \cite{Martin:2009iq}, are considered 
as a cross-check of the PDF-dependence of the  $Z/\gamma^*$ kinematical distributions.

For our studies of the QED FSR effects we use \PHOTOS\ in its best set-up, 
i.e.\ with multi-photon radiation including the ME correction. 
To see how important are  the sub-leading ${\cal O}(\alpha)$ FSR corrections we also perform computations
with the ME correction switched off. 
In order to cross-check the \WINHAC$+$\PHOTOS\ results, 
we use the \KKMC\ Monte Carlo event generator \cite{KKMC:1999}.  
It features the ${\cal O}(\alpha^2)$ YFS exclusive exponentiation of QED radiative
corrections in the process of $Z/\gamma^*$ production and decay. 
Its main advantages with respect to \PHOTOS\ are: 
inclusion of initial-final state radiation interferences, 
the exact  ${\cal O}(\alpha^2)$ FSR corrections 
and a better overall normalisation (includes the radiative corrections, which is not the case in \PHOTOS).   
Initially \KKMC\ was developed for $e^+e^-$ colliders, 
such as LEP, but recently it has been adapted also for hadron colliders by adding quark beams  
and a possibility of their convolution with PDFs. It was already used to assess the size of the second-order
QED correction to the $\phi_{\eta}^*$ observable for the NC DY process at the LHC \cite{Doan:2013qqa}.

Since we want to follow closely the analysis of the LHC experiments, instead of using the QCD $K$-factors,
we normalise the $Z/\gamma ^*$-boson and Higgs-boson cross sections to the values used in the ATLAS analysis \cite{ATLAS_tau}. 

The analysis shown in the current and in the following sections  is made for the $pp$ centre-of-mass 
collision energy of  $8\,$TeV and the event-acceptance cuts: 
\begin{equation}
|\eta_l | \leq 2.5, \qquad
\pTl \geq  10\, \mathrm{GeV}, \qquad
m_{ll} \geq 40\, \mathrm{GeV}, 
\label{eq:cuts}
\end{equation}
where $\eta_l$, $\pTl$ and $m_{ll}$ are, respectively,
the pseudo-rapidity, the transverse momentum of each of the leptons ($\mu^{\pm}$ or $\tau^{\pm}$) 
and the invariant mass of the opposite-sign lepton pairs.

\begin{figure*}
\begin{center}
\includegraphics[width=0.48\textwidth]{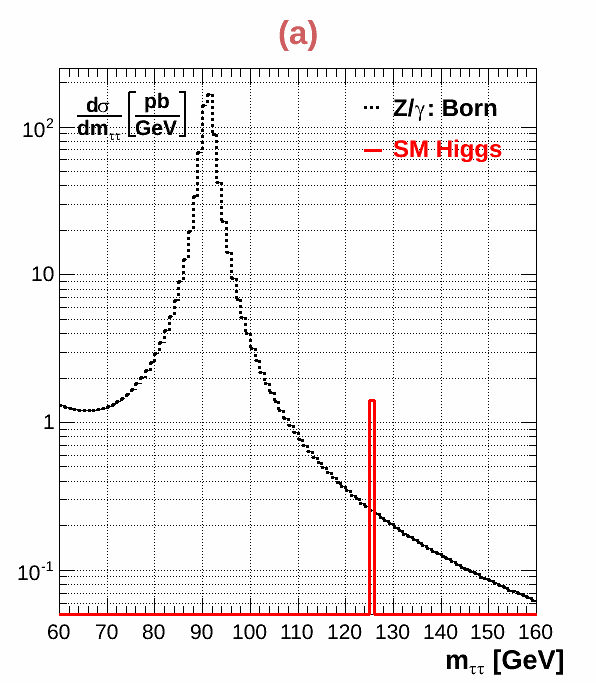} 
\includegraphics[width=0.48\textwidth]{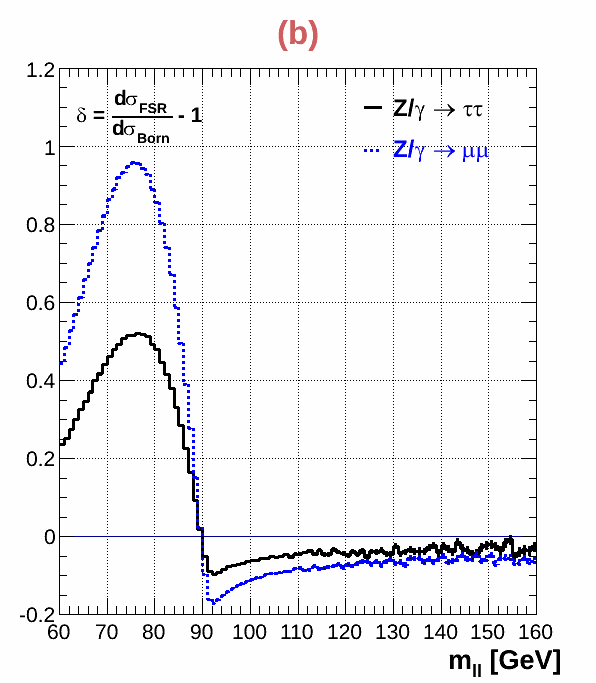} 
\includegraphics[width=0.48\textwidth]{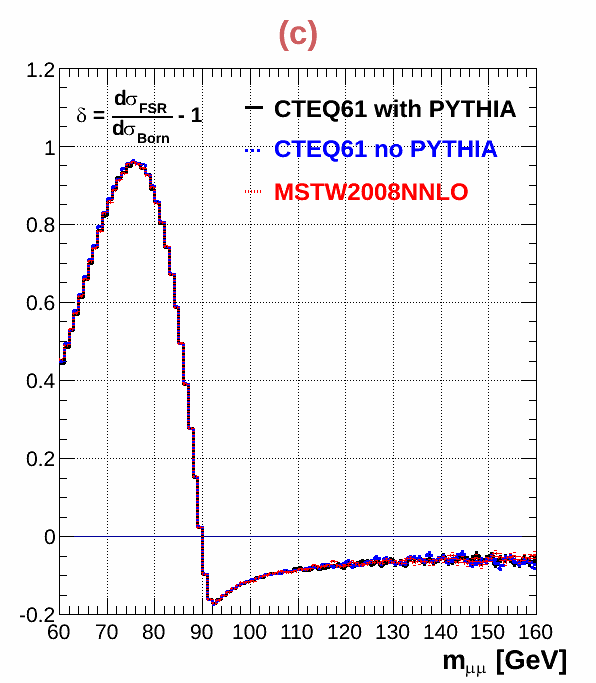} 
\includegraphics[width=0.48\textwidth]{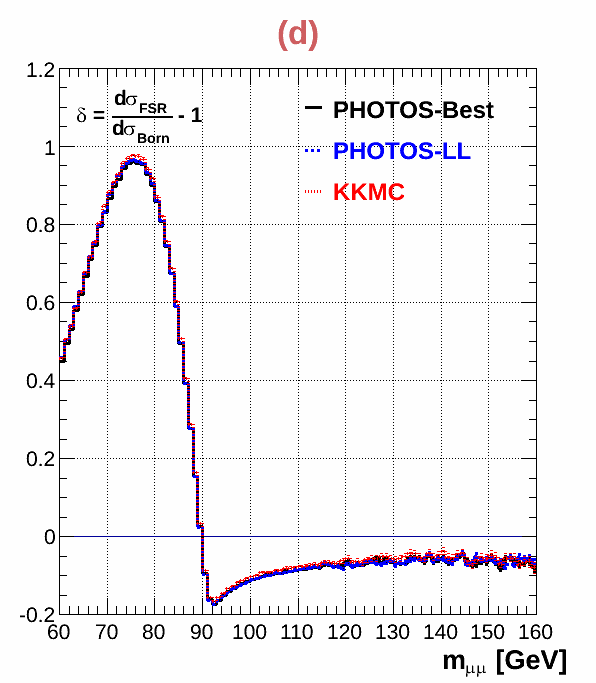} 
\end{center}
\caption{(a) The \WINHAC$+$\PHOTOS\ inclusive opposite sign $\tau\tau$ mass distribution and the expected 
125~GeV Standard Model Higgs signal (shown for a bin width of 1 GeV);
(b) the QED radiative corrections ($\delta = \frac{d\sigma _{\rm FSR}}{d\sigma_{\rm Born} } - 1$) for bare $\mu$ and  $\tau$ pairs;
(c) the PDF and QCD-effects dependence of the size of the QED radiative corrections for the muon pairs (see text for details);
(d) the size of the QED radiative corrections for various calculation approximations (see  text for details).
}
\label{fig1}
\end{figure*}

In Fig.~\ref{fig1} we present the results of the generator-level  precision evaluation of the tools used in this analysis. 
In Fig.~\ref{fig1}a we show the Born-level invariant-mass distribution of the  $\tau ^+ \tau ^-$
pairs which originate from $Z/\gamma ^*$-boson decays and from Higgs-boson decays. 
At the event generator  level the 
Higgs-boson peak is clearly visible above the $Z/\gamma^*$-boson background. 

The QED FSR  corrections ($\delta = \frac{d\sigma _{\rm FSR}}{d\sigma_{\rm Born} } - 1$) to the bare-dilepton invariant mass distribution $m_{ll}$, 
for the $Z/\gamma^*$ decays, are shown in Fig.~\ref{fig1}b, 
separately for $\tau$-pair  and for $\mu$-pair final states. 
As expected, the radiative corrections  for the  $\tau$ case  are significantly ($\sim 2$ times) smaller than 
those for the $\mu$ case.  The FSR corrections are positive below the $Z$-boson mass peak and can be as large as 
$\sim 100\%$ for muons and $\sim 50\%$ for taus. At  $m_{ll} = M_Z$ the corrections  change
sign and become negative.  In the Higgs-boson region ($\sim 125\,$GeV) the corrections are  about $-10\%$ for muons
and about $-5\%$ for taus.

In Fig.~\ref{fig1}c we show that the sensitivity of a realistic simulation of the $Z/\gamma^*$-boson four-momentum for 
$\mu$-pair final states to PDF parametrisations and to  \PYTHIA\ parton-shower and hadronisation effects is small.  
We use {\sf CTEQ6.1} PDFs with and without \PYTHIA, and {\sf MSTW2008NNLO} PDFs with \PYTHIA.
As can be seen, all three curves are  on top of each other.  Clearly,  the results for the FSR corrections
do not depend, within a 1\%, on the QCD patron-shower and hadronisation effects, or  on the choice of the PDFs parametrisation.

Finally, in Fig.~\ref{fig1}d, we present the study of effects of various approximations
used in the calculations of the QED radiative corrections. 
We compare two running modes of \PHOTOS: 
(1) its best set-up, i.e.\ multi-photon radiation including the ME correction, denoted as {\sf PHOTOS-Best},
and (2) the multi-photon radiation without the ME correction, denoted as {\sf PHOTOS-LL}. 
The differences between these two results are very small, below $1\%$, and are not visible in the plot. 
This shows that the ME correction in \PHOTOS, which accounts for the sub-leading ${\cal O}(\alpha)$ 
QED effects, is not important for the  studies presented here. 
Nevertheless, our main results are obtained using \PHOTOS\ in its best set-up. 
In Fig.~\ref{fig1}d, we also compare the \WINHAC$+$\PHOTOS\ radiative corrections with those of \KKMC.  
Here the differences are also barely visible in the plot:
they reach up to $1.5\%$ below the $Z$-boson mass peak and up to $\sim 1\%$ above it. 
These can be attributed to the differences between  \KKMC\ and  \PHOTOS\ discussed earlier.  
We have checked that these differences are almost identical for the $\tau$ and $\mu$ cases.  Therefore, as far as the 
difference of  the radiative corrections for muons and taus is concerned,  
they are negligible. The same is true for the purely weak corrections. 
Therefore, we conclude that \WINHAC$+$\PHOTOS\ is adequate  for our study of
the QED FSR influence on the background to the $H\rightarrow \tau\tau$ signal in 
the LHC Run 1 data. In the future, with increased statistical and systematic precision of the LHC data,
using more precise MC generators, such as \KKMC, may be necessary.  

The above  study demonstrates that at the generator level the radiative corrections for the NC DY process
are well controlled. In the mass region relevant for the Higgs signal they are small compared to the signal
itself. One could thus expect that they would  have  a negligible impact
on the experimental determination of  the $H \rightarrow \tau \tau$ signal at the present statistical precision. 
This is indeed the case for the classical analysis, in which the radiative effects 
are implemented in all the background MC generators and the treatment of 
simulated and data events is the same. 

In the following subsection we demonstrate that this statement 
no longer holds for the embedding-procedure-based analysis which uses real 
rather than generated events.

\subsection{FSR corrections to embedding procedure} 
\label{false-signal}

As discussed earlier,  we use the term {\it bare leptons} for 
those which have their momentum vector  reconstructed solely from the track parameters and  the term {\it dressed leptons} for an idealised case in which all the radiative photons are assigned to their lepton emitters 
and the photon  measured momenta are used in the  reconstruction of the  initial momentum vector 
of the leptons. Therefore the dressed lepton, as far as the FSR is concerned, is equivalent to a Born lepton. 
If in the extraction of the Higgs signal, data are compared to complete Monte Carlo simulations 
of $Z/\gamma ^*  \rightarrow \tau\tau$ and $H  \rightarrow \tau\tau$ events, the analysis would be consistent.
In  the analyses of \cite{ATLAS_tau,CMS_tau}  a fully simulated $H  \rightarrow \tau\tau$ sample is used.
However, the $Z/\gamma ^*  \rightarrow \tau\tau$ sample is obtained by embedding $\tau$-leptons
in a sample of $Z/\gamma ^*  \rightarrow \mu\mu$ data. Had dressed $\tau$-leptons been used 
to replace dressed muons in data,  the analysis would still be consistent. 
However, what is done instead,  is to replace bare muons in data with simulated dressed $\tau$-leptons.

The necessity of taking into account the QED FSR corrections to the embedding 
procedure has its origin in the interplay of the following three  effects:

\begin{itemize}

\item 
in the embedding procedure bare muons in data are replaced by simulated dressed $\tau$-leptons,

\item
the measurement resolution of the invariant mass of two opposite-charge $\tau$-leptons ($\tau ^+ \tau ^-$)
is comparable to  the mass difference  between the $Z$-boson and the Higgs boson,

\item 
the inclusive cross section for $Z/\gamma^*$-bosons  decaying into leptons is  three orders of magnitude
larger  than that for the Higgs boson. 

\end{itemize}

\begin{figure*}
\begin{center}
\includegraphics[width=0.89\textwidth]{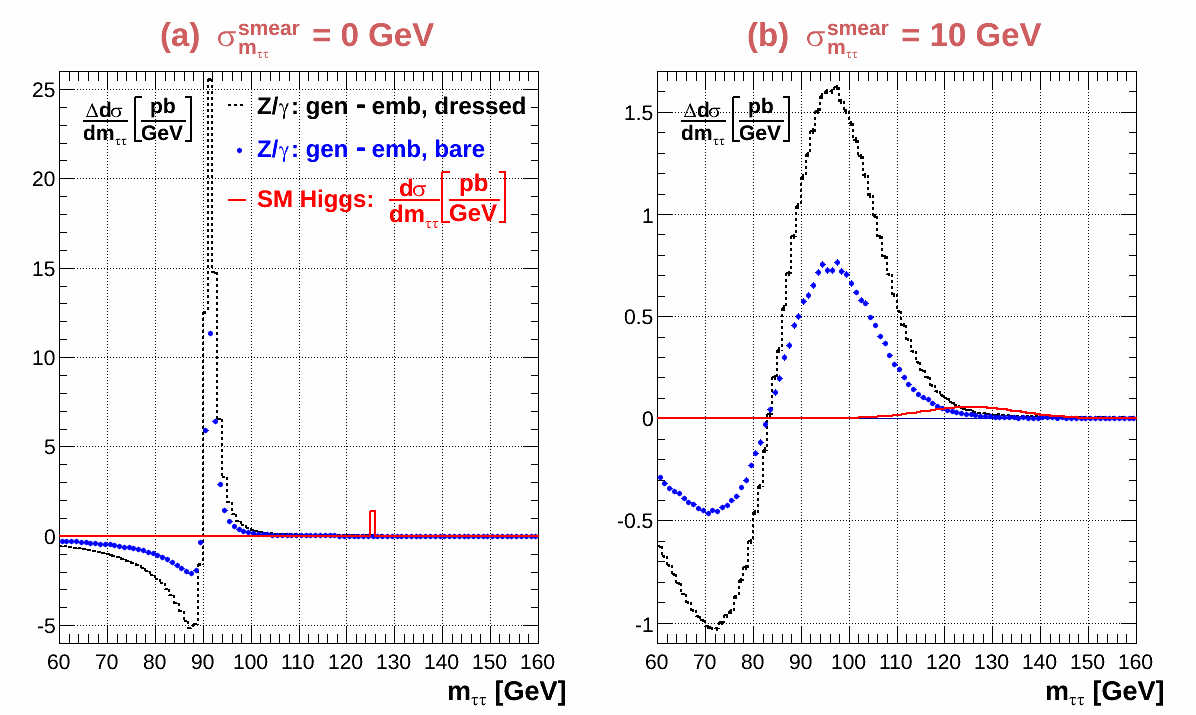} 
\includegraphics[width=0.89\textwidth]{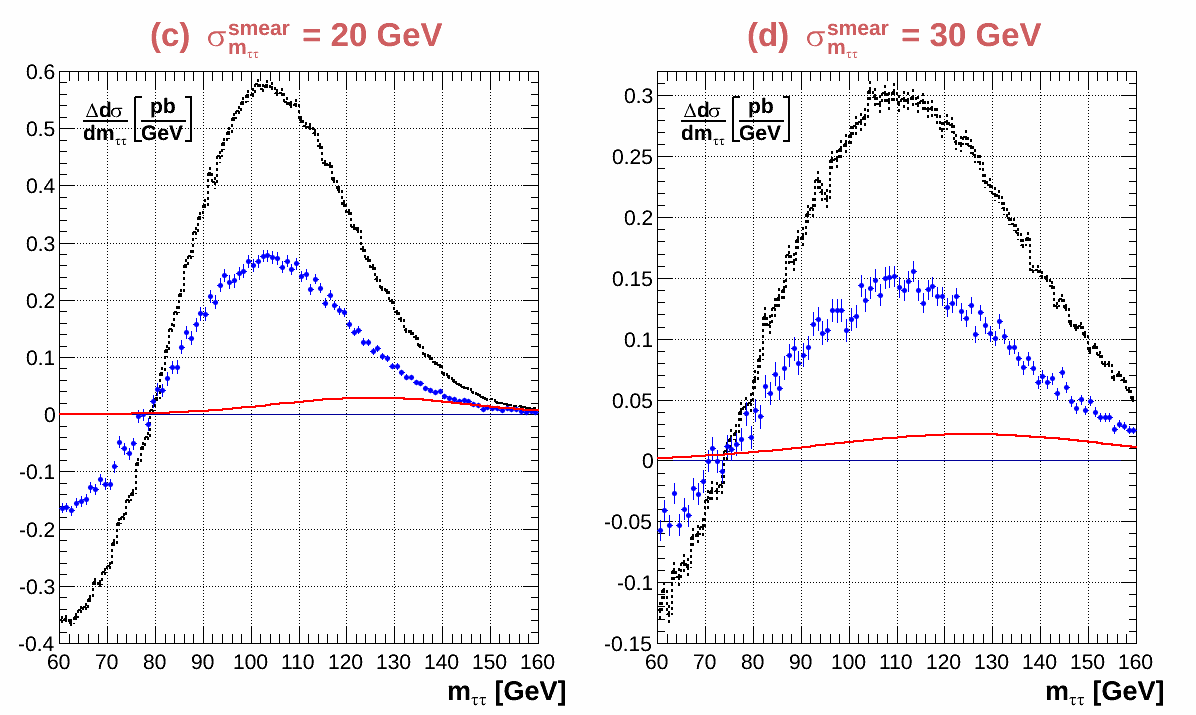} 
\end{center}
\caption{The dressed and bare QED radiative corrections to the embedding procedure,
$  \frac {\Delta d \sigma}{dm_{\tau\tau}} = \frac{d\sigma_{\mathrm{generated}}}{dm_{\tau\tau}} - \frac{d\sigma_{\mathrm{embedded}}}{dm_{\tau\tau}}$, 
for $pp$ centre-of-mass collision energy of  $8\,$TeV and the following event-acceptance cuts: 
$|\eta_l | \leq 2.5$, 
$\pTl \geq  10\, \mathrm{GeV}$ and 
$m_{ll} \geq 40\, \mathrm{GeV}$. 
Shown are plots for the following four values of $\sigma ^{\rm smear} _{m_{\tau\tau}}$  specifying 
the Gaussian  smearing of the measured invariant mass of the $\tau ^+ \tau ^-$
pairs: $\sigma ^{\rm smear} _{m_{\tau\tau}}$ = 0, 10, 20 and 30 GeV -- shown 
in panels  (a), (b), (c) and (d), respectively. The corresponding smeared  distributions for the Standard Model (SM) Higgs-boson signal,
$\frac {d \sigma}{dm_{\tau\tau}}$, are also shown for comparison.}
\label{fig2}
\end{figure*}

The interplay of the above three effects is illustrated in Fig.~\ref{fig2}.
The upper left plot (a), shows the difference between the \WINHAC$+$\PHOTOS\ differential cross section 
plotted as a function of the invariant mass of the $\tau ^+ \tau ^-$ pair, 
$m_{\tau\tau}$, for the generated sample of the $Z/\gamma ^*  \rightarrow \tau\tau$ decays and for the 
corresponding cross section for the case,  where  the $\tau$-lepton momenta are 
determined using the embedded $Z/\gamma ^* \rightarrow \mu\mu $ sample. 
We call the differential cross-section difference:
\begin{eqnarray}
 \frac {\Delta d \sigma}{dm_{\tau\tau}} = \frac{d\sigma_{\mathrm{generated}}}{dm_{\tau\tau}} - \frac{d\sigma_{\mathrm{embedded}}}{dm_{\tau\tau}},
\end{eqnarray}
{\it the radiative corrections to the embedding procedure}. 
Note, that these corrections are additive rather than multiplicative.
Such a representation is appropriate for  the studies of  radiative corrections which are ``non-local",  i.e. 
for  the case when their size in  each mass bin  is an  integral over the  Born cross section  
for  a wide  mass interval.

The procedure of association of the radiative photons with the parent 
$\tau$-leptons in the reconstruction is  not discussed in \cite{ATLAS_tau,CMS_tau}.
Therefore,  we cannot provide  in this paper  a precise estimate of 
the radiative corrections to the embedding procedure.  Instead, 
we estimate  upper and  lower boundaries of the radiative corrections. 
Below we refer to these two boundaries as (1) {\it dressed corrections},  and (2) {\it bare  corrections}.

The two distributions shown in Fig.~\ref{fig2}a reflect the following  two possible  outcomes 
of the reconstruction of the outgoing $\tau$-lepton momenta 
in  generated $Z/\gamma ^* \rightarrow \tau\tau$ decays: 
(1) the momenta of the dressed $\tau$-leptons -- represented by 
the dashed line (dressed corrections) and (2) the momenta of the bare  $\tau$-leptons -- represented  
by the dots (bare  corrections). 
These correspond to boundaries on the size of the radiative corrections to the embedding procedure. 

If, in the event-by-event experimental reconstruction procedure of the $\tau$-lepton momenta,  
all photons radiated by each of the $\tau$-leptons were reconstructed from the electromagnetic 
energy clusters in the calorimeters and if they were unambiguously associated to one of the 
outgoing $\tau$-leptons, contributing  to its momentum reconstruction,  the radiative 
corrections would be those represented by the dashed line. If, on the other hand, all the 
calorimetric energy depositions of photons coming from the initial $\tau$-lepton radiation 
(before its decay) were  not associated to their parent  $\tau$-leptons, the radiative 
corrections to the embedding procedure would have the values represented by the dots.

The solid line in Fig.~\ref{fig2}a represents the cross section for the 
$H \rightarrow \tau \tau$ decay signal for the case of perfect resolution.
It is significantly higher than the size
of the radiative corrections. Thus, are the radiative corrections 
irrelevant while establishing the existence of the Higgs-boson signal? 
If the invariant mass of the $\tau$-lepton pairs, $m_{\tau\tau}$, 
could be  reconstructed with the precision comparable to that for the muon  pairs, 
 the answer would be affirmative and the discussion in this chapter could be considered as purely academic.
In reality, however,  the mass of the $\tau$-lepton pairs is 
reconstructed with  finite resolution, insufficient to assign  
the $\tau$-lepton pairs to either  the $Z/\gamma^*$-boson or  Higgs-boson source, on an event-by-event basis. 

In Fig.~\ref{fig2}b--d we show  the size of the radiative correction to the embedding procedure
for the following three values of $\sigma ^{\rm smear} _{m_{\tau\tau}}$  specifying the Gaussian smearing of 
the reconstructed $m_{\tau\tau}$: $\sigma ^{\rm smear} _{m_{\tau\tau}}$ = 10, 20 and 30 GeV.
Already for the $\sigma ^{\rm smear} _{m_{\tau\tau}} = 10$ GeV smearing,  the values  of the radiative corrections 
in the mass region of 110--140~GeV 
are comparable to the Higgs-boson signal. The size of the corrections increases rapidly, 
in the Higgs signal mass region,  with worsening resolution
(increasing  $\sigma ^{\rm smear} _{m_{\tau\tau}}$), outweighing the Higgs-boson signal already for 
$\sigma ^{\rm smear} _{m_{\tau\tau}}$ = 20~GeV. We recall that these plots were made  
for the inclusive event sample specified by the kinematical cuts defined in the previous section.
It is evident  that, for any realistic detector and,  in particular,  in the presence of 
escaping neutrinos, these corrections must be considered with  great care. 
 
 It is also  important to note  that the radiative corrections to the embedding procedure 
 have a positive sign in the mass region above the $Z$-boson 
 mass peak and a negative sign below it. If not taken into account, the $Z/\gamma ^* \rightarrow \tau\tau$ background
 to the Higgs searches,  determined using the embedding procedure,  would  be significantly (with respect to the Higgs-boson
 signal strength) underestimated in the Higgs-boson mass  search region 
 above  $Z$-boson mass and overestimated for  masses smaller than the $Z$-boson mass.
 It is thus precisely in the region where $H \rightarrow \tau \tau$  decay signals have been reported \cite{ATLAS_tau,CMS_tau} that
 the neglected radiative corrections to the embedding procedure may potentially produce a spurious Higgs-boson signal
 by underestimating  the background level.

\section{$H \rightarrow \tau\tau$ evidence}
\label{bias}

\subsection{Initial remarks}
\label{iremarks}

The ATLAS and CMS papers discussing  the  $H \rightarrow \tau \tau$  decay evidence \cite{ATLAS_tau,CMS_tau} 
do not present unfolded signal and background distributions in which the detector and analysis 
dependent corrections are accounted for. The results presented in such a form 
cannot, strictly speaking,  be fully checked by any external analysis. 
The goal of  the analysis presented in this section is only to provide the best possible {\it initial estimate} 
of the size  of the missing radiative corrections in the modelling
of the $Z/\gamma^* \rightarrow \tau\tau$  background and an initial assessment  of their 
impact on the extraction of the  $H \rightarrow \tau \tau$  decay signal.  Since several simplifications  have to be made, 
we  specify the details of these estimates, leaving to the LHC collaborations the precise
determination of the size of the corrections for their respective  analysis  methods. 

As discussed in the previous section, the effective size of the missing radiative corrections 
in the modelling
of the $Z/\gamma^* \rightarrow \tau\tau$  background 
for the ATLAS and CMS analyses are sensitive to the experimental resolution of the measurement of 
 $m_{\tau\tau}$. 
The corrections also depend on  the efficiencies of the event-subsample  selection cuts,  which were  
optimised in the ATLAS and CMS analyses  \cite{ATLAS_tau,CMS_tau}
to enrich the $H \rightarrow \tau \tau$  signal   with respect to all background contributions.

The event-subsamples in the $H \rightarrow \tau \tau$  searches are classified in terms of  event categories. 
These categories are characterised by: (1) the type of the $\tau$-lepton decays:  $\tau _e(\tau_{\mu}) \tau_{had}$,
 $\tau _e \tau_{\mu}$, $\tau_{had} \tau_{had}$, and (2)  the event jet activity and the jet(s) transverse momentum.
CMS considers 38 event categories,  while  ATLAS considers 6 categories (more information on the event categories 
can be found in Ref.  \cite{ATLAS_tau,CMS_tau}). In each of the event categories the $m_{\tau\tau}$  resolution and the  
ratio of the signal to background  are different.

\subsection{The CMS case}
\label{CMS}

\begin{figure*}
\hspace{3.5cm}
(a)
\hspace{7.5cm}
(b)
\begin{center}
\includegraphics[width=0.35\textwidth]{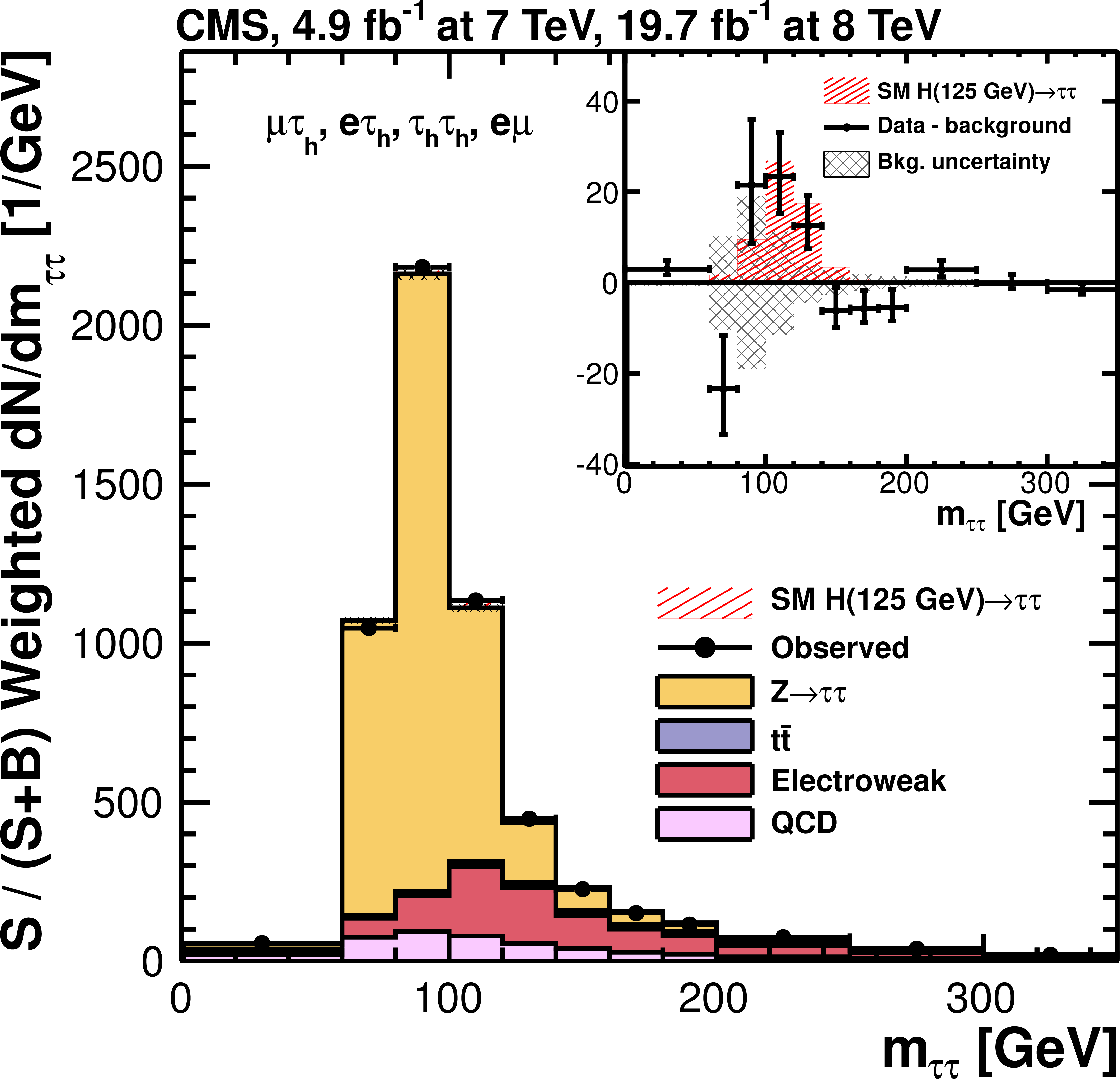} 
\includegraphics[width=0.63\textwidth]{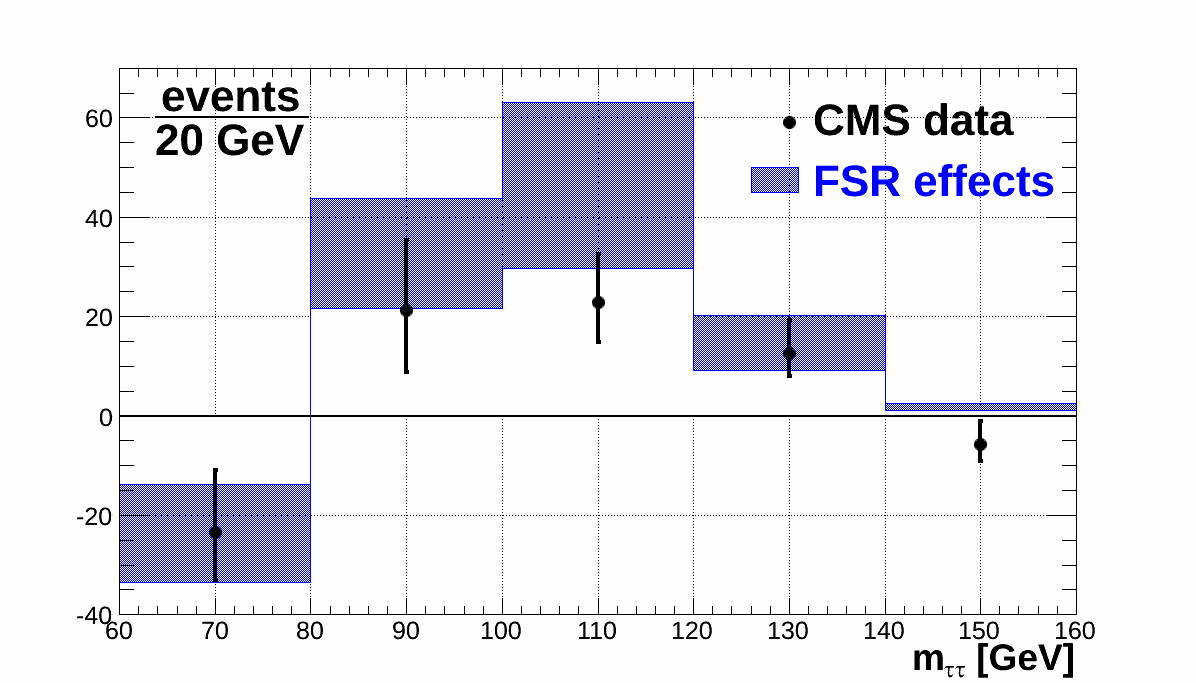} 
\end{center}
\caption{(a) The CMS Fig.~11 taken from \cite{CMS_tau};
(b) the comparison of the observed excess of weighted events, interpreted in  \cite{CMS_tau} as the Higgs signal, with
our predictions 
for the excess of weighted  background events  that originate from the  missing FSR corrections to the embedding procedure
of modelling the $Z/\gamma^* \rightarrow \tau\tau$  background.}
\label{fig3}
\end{figure*}

In Fig.~\ref{fig3}a we reproduce Fig.~11 of the CMS paper  \cite{CMS_tau}. Fig.~\ref{fig3}a  shows the observed 
and predicted $m_{\tau\tau}$ distribution integrated over all the CMS event categories. 
The normalisation of the predicted background distributions corresponds to the result of a global fit.
The Higgs-signal distribution is normalised to the Standard Model (SM) prediction ($\mu$ = 1).
The distributions in each category, $i$,  are weighted by the ratio between the expected signal and ``signal+background" 
yields, $w_i = S_i/(S_i +B_i) $.
The insert shows the corresponding difference 
between the data and their estimate of the background distributions.  Also
shown is the expected  distribution 
for a 125 GeV Standard Model Higgs boson.

We estimate the effect  of the missing radiative corrections 
 to the embedding procedure in the modelling
of the $Z/\gamma^* \rightarrow \tau\tau$  background
 in three steps.

In the first step we determine, from Fig.~\ref{fig3}a,  the total number of weighted  $Z/\gamma^* \rightarrow \tau\tau$ events in the mass interval: 60--160~GeV to be  $N^w_{tot} =  3780 \pm 190$. 
In the second step we determine the Gaussian smearing parameter 
$\sigma ^{\rm smear} _{m_{\tau\tau}}(\rm CMS)$ 
of the  $m_{\tau\tau}$ measurement as the average of the  
38 RMS values, $\sigma _i $,  for each of the  $\rm CMS$ event categories:  
$ \sigma ^{\rm smear} _{m_{\tau\tau}}(\rm CMS) = \langle \sigma _{i} \rangle = 18.4$ GeV. 
The values for each of the event categories are taken from Table 4 in  Appendix~B of Ref.  \cite{CMS_tau}.
In the third step we calculate our predictions for the QED radiative corrections 
to the  $Z/\gamma^* \rightarrow \tau \tau$    background contribution shown in  Fig.~\ref{fig3}a. 

We  first calculate the  difference between the differential cross section
for the generated sample of $Z/\gamma^* \rightarrow \tau\tau$ decays and the 
corresponding cross section in which the $\tau$-lepton momenta are 
determined by using the embedded $Z/\gamma^* \rightarrow \mu\mu $ sample. 
 We do it twice:  for the bare and for the 
dressed lepton corrections  with  the Gaussian mass smearing 
of  $ \sigma ^{\rm smear} _{m_{\tau\tau}}(\rm CMS)  = 18.4$ GeV.
The cross sections are  subsequently multiplied by the ``weighted"  luminosity 
defined such that  the total number of the 
embedded $Z/\gamma^* \rightarrow \tau\tau$ events in  the mass internal 60--160~GeV 
is equal to the number of the weighed events in the CMS data, $N^w_{tot}$. 
This normalisation allows us to replace the calculation of the radiative corrections to the 
differential cross sections by the calculation of  the number 
of weighted events  which must be added to the $Z/\gamma^* \rightarrow \tau\tau$ background 
contribution, shown in Fig.~\ref{fig3}a, in order to take into account the missing radiative corrections
in the modelling of the $Z/\gamma^* \rightarrow \tau\tau$  background.

Such a  procedure involves the following three  simplifications:
\begin{enumerate} 
\item
The $m_{\tau\tau}$ smearing is assumed to be Gaussian. 
\item
The small effect of the radiative corrections to the  category dependent weights, 
$w_i$, which requires their redefinition as $w_i^{'} = S_i/(S_i +B_i(1 + \delta _i)) $,
where $\delta _i$ are the category-dependent radiative corrections,  is neglected. 
\item
Instead of calculating 
the radiative correction for each of the event categories,  with the category-dependent 
$  \sigma _i$,  we use its  average value $\sigma ^{\rm smear} _{m_{\tau\tau}}(\rm CMS)$.
The latter simplification assumes implicitly that: 
\begin{eqnarray}
 \sum_{i=1}^{38} \sigma _i  w_i   N_i  \approx \sigma ^{\rm smear} _{m_{\tau\tau}}({\rm CMS})  \sum_{i=1}^{38} w_i  N_i,   
 \end{eqnarray} 
where $i$ denotes the event category and $N_i$ the category-dependent number of 
$Z/\gamma^* \rightarrow \tau\tau$ background events satisfying the CMS event-selection criteria. 
This condition is strictly fulfilled only in  the limit of a negligible dispersion of the  $  \sigma _i$  values.
In reality the dispersion of  the $  \sigma _i$  values is 3.1~GeV.  This value is 
sufficiently small to justify such a simplification.  
In order to quantify such a statement   we repeated  the calculations for  $  \sigma ^{\rm smear} _{m_{\tau\tau}} = 15$~GeV  and 
for $  \sigma ^{\rm smear} _{m_{\tau\tau}} = 20$~GeV. The corresponding radiative corrections are found to stay well 
within  the  uncertainty range  of the radiative corrections which spans the difference  between the bare and dressed 
radiative corrections. 
\end{enumerate}

In Fig.~\ref{fig3}b we compare our predicted
excess of weighted events that originate from 
 the missing QED radiative corrections in the modelling
of the $Z/\gamma^* \rightarrow \tau\tau$  background
   with the observed excess 
of weighted events, taken from the insert of the plot Fig.~\ref{fig3}a, and 
interpreted in  Ref. \cite{CMS_tau} as  the Higgs-boson 
signal. Our predictions are plotted as the shadowed regions, the upper limit corresponding 
to the dressed radiative corrections and the lower limit  corresponding to the bare corrections. 
As can be deduced from this plot, the missing radiative corrections in the modelling
of the $Z/\gamma^* \rightarrow \tau\tau$  background 
 could  fully account for 
the excess of events and the Higgs-boson contribution may  no longer be required. 
The $p$-value qualifying the agreement of the predictions of the excess of events 
due to the missing radiative corrections in the modelling
of the $Z/\gamma^* \rightarrow \tau\tau$  background with the data is $0.55$. It is  calculated  
for  the arithmetic mean of the predictions for the bare and for the dressed 
radiative corrections with the prediction error taken as a half of the difference of these predictions. 
The corresponding $p$-value for the hypothesis in which the Higgs signal is added on top of the missing radiative corrections 
in the modelling of the $Z/\gamma^* \rightarrow \tau\tau$  background is $0.007$. Such a hypothesis is clearly disfavoured by the data.

As discussed in Subsection \ref{false-signal}, a  striking footprint of the radiative-corrections-driven 
excess of events  is that it  changes sign for $m_{\tau\tau}<M_Z$.
 This  effect,  although statistically weak at the present collected luminosity,  is indeed observed 
 in the data and well reproduced by the missing radiative corrections in the modelling
of the $Z/\gamma^* \rightarrow \tau\tau$  background, thus supporting
 the  hypothesis that the excess events may  be fully accounted for by 
 the effects of the  missing radiative corrections.  

\subsection{The ATLAS  case}
\label{ATLAS}

Our initial estimate of the effect of the missing radiative correction to the embedding procedure
in the modelling of the $Z/\gamma^* \rightarrow \tau\tau$  background
for the ATLAS analysis presented in Ref.  \cite{ATLAS_tau} is made separately for the six event categories.

 The ATLAS  event classes reflect the decay channels   of the $\tau$-leptons: $\tau_{lep}\tau_{lep}$, 
 $\tau_{had}\tau_{lep}$ and $\tau_{had}\tau_{had}$. For each of the above three classes,  
 two hadronic remnant configurations, assuring an optimal rejection efficiency of  background  events 
 with respect to the Higgs-boson signal,  are defined: (1) the {\em Vector-Boson-Fusion (VBF)}
 configuration,  in which the production of the $\tau$-lepton pair is accompanied by 
 the emission of two jets having a large separation in rapidity, (2) the {\em Boosted} 
 configuration in which the $\tau$-lepton pair  recoils against a high transverse-momentum jet(s).
 The above three decay channels  and the two  hadronic configurations define six event categories.
Additional details about the  selection criteria of events belonging to each of the above categories
 can be found in Ref.  \cite{ATLAS_tau}.

In our initial  estimate of the radiative correction effects to the CMS analysis we assume Gaussian 
smearing of the reconstructed $m_{\tau\tau}$. For the ATLAS case we base our analysis 
on the published resolution functions presented in Ref. \cite{ATLAS_tau} for the VBF and Boosted categories.
This allowed us to take into account non-Gaussian resolution tails. 

\begin{figure*}
\hspace{4.2cm}
(a)
\hspace{8.0cm}
(b)
\begin{center}
\includegraphics[width=0.49\textwidth]{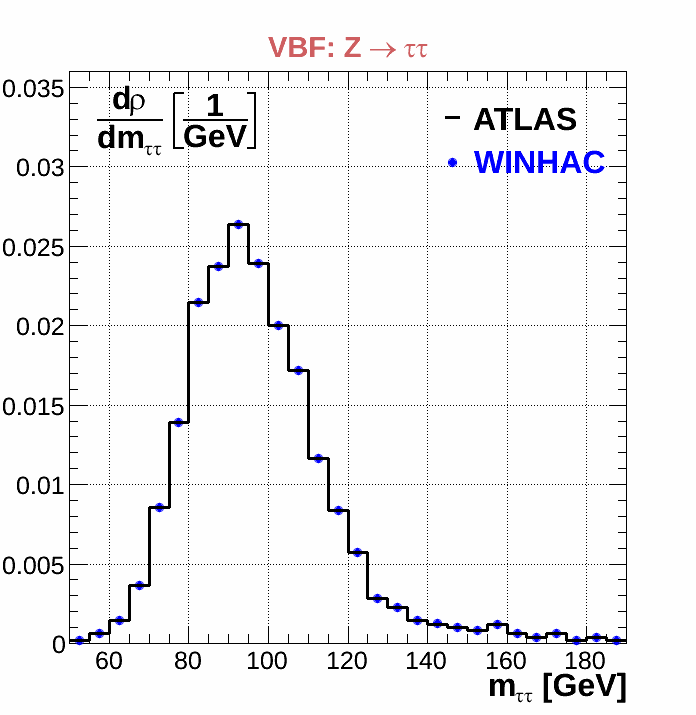} 
\includegraphics[width=0.49\textwidth]{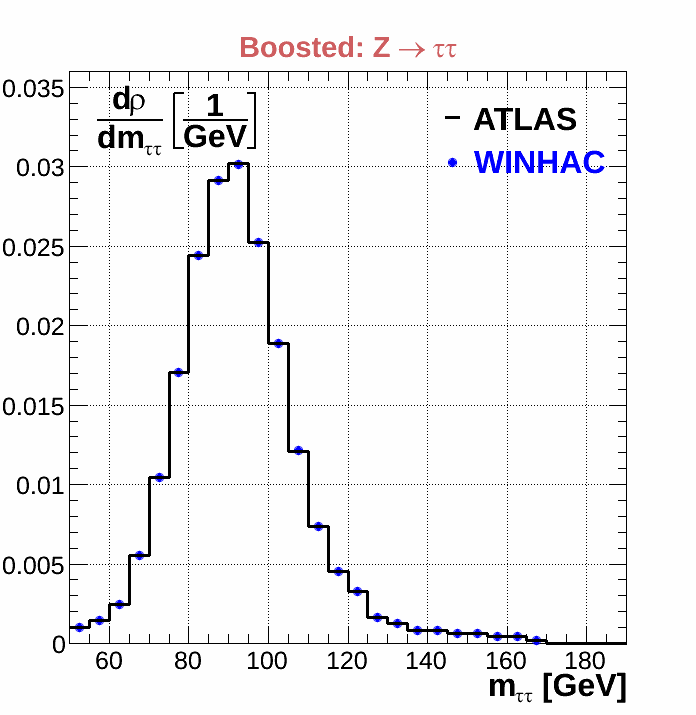} 
\end{center}
\caption{The reconstructed invariant $\tau ^+ \tau ^-$  mass, $m_{\tau\tau}$, 
in the ATLAS experiment  \cite{ATLAS_tau} compared with the \WINHAC-based
model used in our studies: (a)  for the VBF selection of events,  and  (b) for the Boosted selection 
of events.}
\label{fig4}
\end{figure*}

In Fig.~\ref{fig4}a and  Fig.~\ref{fig4}b we compare the ATLAS  $m_{\tau\tau}$ distributions
for $Z/\gamma^* \rightarrow \tau \tau$  decays, for the VBF and Boosted configurations, 
with the $m_{\tau\tau}$  distribution based on our model of the measurement resolution.  
The published and our model distributions are in  very good agreement 
which validates our modelling of the experimental resolutions. 
All the results presented below are be based on  this modelling of the experimental resolution.
The resolution functions are assumed  to be the same for the $\tau_{lep}\tau_{lep}$, 
 $\tau_{had}\tau_{lep}$ and $\tau_{had}\tau_{had}$ decay modes.

\begin{figure*}
\hspace{2.7cm}
(a)
\hspace{8.0cm}
(b)
\begin{center}
\includegraphics[width=0.30\textwidth]{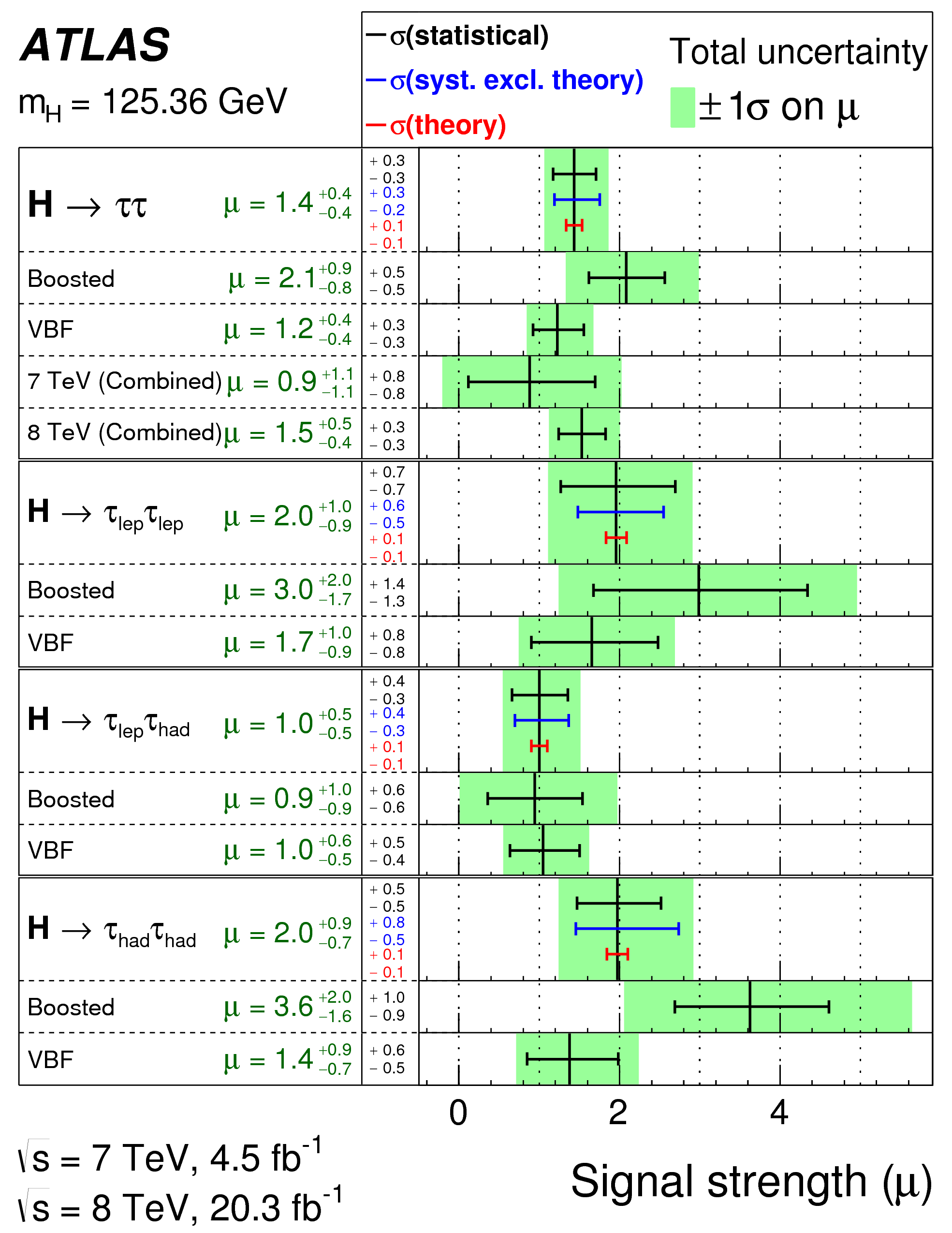} 
\includegraphics[width=0.68\textwidth]{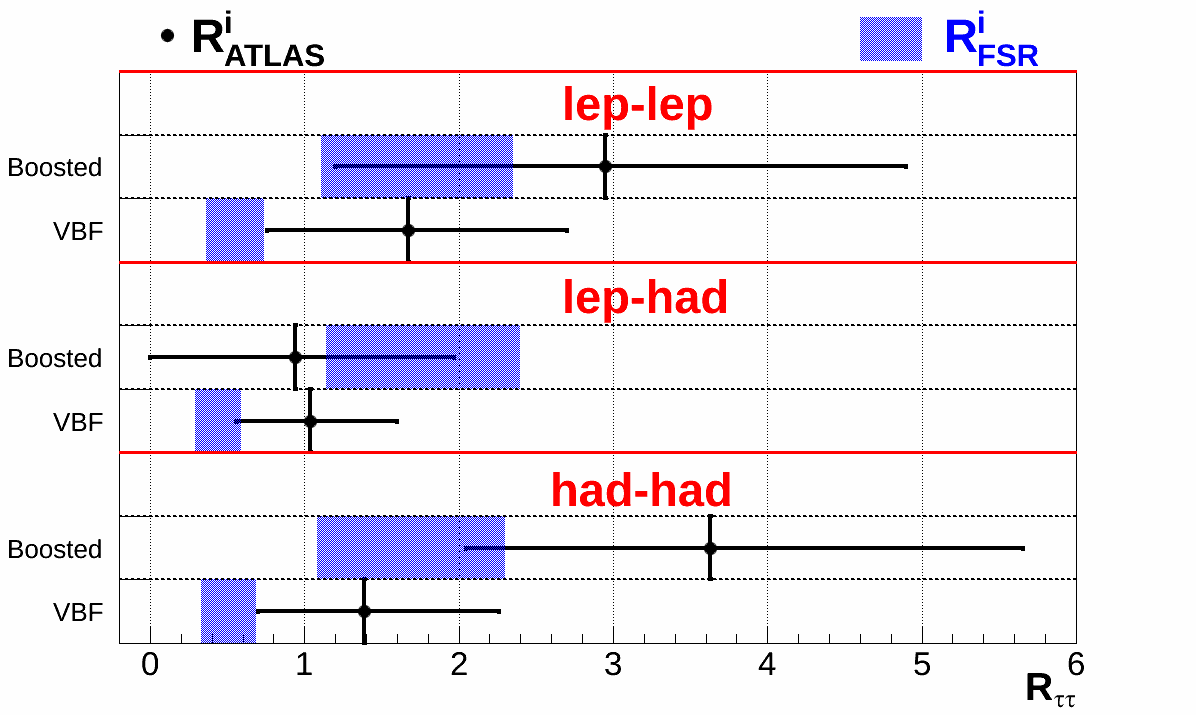} 
\end{center}
\caption{(a) Reproduction of  Fig.~10 taken from the ATLAS paper\cite{ATLAS_tau};
(b) a comparison of the $R_{\rm ATLAS}^i$ values  taken from Fig.~10 of Ref.  \cite{ATLAS_tau},
with the $R_{\rm FSR}^i$ values for each event category  
(here, the upper boundaries of the shadowed regions for the $R_{\rm FSR}^i$ values correspond to   
 dressed corrections and the lower boundaries correspond to  bare corrections).}
\label{fig5}
\end{figure*}

The ATLAS selection criteria for the Boosted and VBF categories increase the 
ratio of the number of  accepted $H \rightarrow \tau\tau$ events  sizeably with respect  to 
the $Z/\gamma^* \rightarrow \tau\tau$ background,   as compared  to the fully inclusive samples. 
The ratios of the integrated number of the accepted $H \rightarrow \tau\tau$ events, $N_{\rm tot}^{H,i}$,   to the number of the 
accepted $Z/\gamma^* \rightarrow \tau\tau$ events, $N_{\rm tot}^{Z,i}$, are determined in our analysis 
for each event category
using  Fig.~2 of Ref.  \cite{ATLAS_tau}. These ratios are used 
to calculate  the corresponding ratios of the event-selection efficiencies in the ATLAS-analysis 
acceptance region  \cite{ATLAS_tau}:
\begin{equation}
 \frac{\epsilon _H^i}{\epsilon _Z^i} = \frac{N_{\rm tot}^{H,i}}{N_{\rm tot}^{Z,i}} \times  \frac{\sigma _Z \times BR(Z \rightarrow \tau\tau)}{\sigma _H \times BR(H \rightarrow \tau\tau)}\,,
\label{eq:ATLASeff} 
\end{equation}
where   superscript $i$ denotes the event category. 
The values of $\sigma _Z \times BR(Z  \rightarrow \tau\tau)$ and $\sigma _H  \times BR(H \rightarrow \tau\tau)$
 taken from \cite{ATLAS_tau}  are  $1303$~pb and $1.4$~pb, respectively. 
 The ratios of the event selection efficiencies are used in turn  to calculate the corresponding ratios of the number of 
 excess events that originate from the  missing radiative corrections
in the modelling of the $Z/\gamma^* \rightarrow \tau\tau$  background,  to the number of 
 expected excess  events from $125$~GeV SM Higgs-boson decays in the ATLAS acceptance regions. 
 These ratios,  calculated for the  $m_{\tau\tau}$ region between 80 and 150 GeV, and denoted  as $R_{\rm FSR}^i = N_{\rm FSR}^i / N _H^i$, can be directly compared to the values of the event-category-dependent $\mu ^i$ parameters,  which are the ratios of the observed to predicted 
 Higgs-boson coupling strength to $\tau$-leptons  in each event category \cite{ATLAS_tau}.  
 This comparison is shown in Fig.~\ref{fig5}.
 
 In Fig.~\ref{fig5}a we 
 reproduce  Fig.~10 taken from the published ATLAS paper \cite{ATLAS_tau},
which shows the $\mu ^i $ values for each event category. These values can be interpreted as 
 the event-category-dependent ratios of the number of  the observed excess events, with respect to the 
 SM background predictions, $N _{\rm excess}^i$,  to the respective number of events originating from 
 the Higgs-boson decays: 
 $R_{\rm ATLAS}^i = N _{\rm excess}^i  /N_H^i$, where, as before,  $i$ specifies the event category.
 If the excess of the observed events is produced only by  $H \rightarrow \tau\tau$ decays, then  
 $R_{\rm ATLAS}^i = \mu ^i = 1$. We refer to this  case as {\it the Higgs-only hypothesis}. 
 If the observed excess of events  is fully accounted for by the missing radiative corrections in the modelling
of the $Z/\gamma^* \rightarrow \tau\tau$  background, 
 $N_{\rm excess}^i =  N_{\rm FSR}^i$,  then $R_{\rm ATLAS}^i = R_{\rm FSR}^i$. We refer to this case  
 as the  {\it the FSR-only hypothesis}. Finally, if the observed excess of events 
 is accounted for by contributions from  both mechanisms,  $R_{\rm ATLAS}^i = R_{\rm FSR}^i +1$. We refer to this case 
 as the {\it the FSR+Higgs hypothesis}.

In Fig.~\ref{fig5}b we compare the values of $R_{\rm ATLAS}^i$  taken from Fig.~\ref{fig5}a
with the $R_{\rm FSR}^i$ values, for each event category. 
The upper boundaries of the shadowed regions for the $R_{\rm FSR}^i$ values correspond to   
dressed corrections and the lower boundaries correspond to  bare corrections.

The first observation is that the FSR-only hypothesis  provides a 
satisfactory description of the data -- the discrepancy with the data is always less than 1$\sigma$ 
and the $p$-value for this hypothesis is $0.46$.  Again, as in the CMS case,
$R_{\rm FSR}^i$ is  calculated  for  the arithmetic mean of the predictions for the bare and dressed 
radiative corrections with the prediction error taken as a half of the difference of these predictions. 
Contrary to the CMS case, for ATLAS  the FSR+Higgs hypothesis 
is  the most  probable. The $p$-value for this 
hypothesis is $0.77$, slightly better than for the FSR-only hypothesis. 
Therefore, although the ATLAS data does not require a contribution from Higgs-boson decays
to explain the observed excess of events, a contribution from Higgs-boson decay cannot
be ruled out given the level of  current  experimental uncertainties.

It is intriguing that our  prediction for the excess of events originating from the  missing radiative corrections 
in the modelling of the $Z/\gamma^* \rightarrow \tau\tau$  background
for the preselected VBF samples is about a factor of 2  lower than the prediction for the Boosted samples.
A similar effect (at a level of 1$\sigma$) 
is observed in the ATLAS  data. The value of  $R_{\rm ATLAS}^i$ for the VBF  category is 
also about a factor of 2 lower than  that for the Boosted category. 
This effect calls for  a further comment. 

Our predictions of the radiative corrections to the embedding procedure for the ATLAS case,
in particular the ratio of the size of the effect for the VBF and Boosted categories, 
rely fully on the ATLAS-analysis-specific efficiencies for the selection of  
Higgs-decay events in the Monte Carlo signal sample and for the selection of   $Z/\gamma ^*$-boson-decay events 
in the embedded data sample. In the ATLAS analysis the Monte Carlo events for  
the VBF Higgs production category are generated with the {\sf POWHEG}$+${\sf PYTHIA8} generators
\cite{POWHEG,PYTHIA8}.  As shown in an earlier ATLAS paper \cite{VBF-Z_ATLAS},
devoted to the VBF $Z/\gamma ^*$-boson production process, both the {\sf POWHEG} and  {\sf SHERPA} \cite{SHERPA}
generators fail to describe the data in the phase-space region \cite{VBF-Z_ATLAS} which is  relevant for the analysis
presented here.
Namely, the Monte Carlo predictions are too high  by a factor of $\sim$2 in the region of large
two-jet invariant mass (see Fig.~11 in  \cite{VBF-Z_ATLAS}) and should be corrected
to obtain the  background templates used in extraction of the $Z/\gamma ^*$-boson VBF signal.
Such corrections may also be important for  adequate modeling of the Higgs-boson VBF signal.
Their absence  in the Higgs-boson signal analysis presented in Ref. \cite{ATLAS_tau}
may explain the apparent discrepancy  of  the  $R_{\rm ATLAS}^i$ and $R_{\rm FSR}^i$
values for the VBF and Boosted categories.
If both the Higgs-boson  and the $Z/\gamma ^*$-boson contributions were to be  taken from Monte Carlo
the impact of these  corrections on  the ratio of efficiencies for the selection of Higgs decay events
and $Z/\gamma ^*$-boson decay events,  would be significantly smaller than in the case
of  the embedding procedure, and the observed discrepancy may disappear.

\subsection{Final remarks}
\label{fremarks}

The results presented in this section represent the initial estimation of the 
size of the radiative corrections to the embedding procedure for  the full acceptance 
regions of the ATLAS and CMS analyses. No attempt is made to determine
how the effective corrections would change in more restrictive kinematical regions or 
for  the Boosted Decision Tree (BDT) based analysis,  in which all the kinematical properties of
the background and signal events are taken into account. 
Since the BDT analyses of ATLAS and CMS  \cite{ATLAS_tau,CMS_tau}
are  based on  distributions which are  not corrected for detector effects, we are technically 
unable to provide the radiative corrections for the BDT-type of analyses.
Our analysis shows, however, that   the ATLAS and CMS  BDT analyses should 
have taken into account the modifications of the templates of the kinematical distributions
for the $Z/\gamma^* \rightarrow \tau\tau$ samples caused by the QED radiative 
corrections to the embedding procedure. This statement holds for 
the present statistical accuracy of the measurement. For a sizeably larger
collected luminosity it may, in principle,  be possible to reduce the size of the 
radiative correction effects by performing the analyses in more severely  restricted phase-space 
regions,  in which the impact of radiative corrections is diminished.

\section{Embedding procedures and FSR effects}
\label{overlooked}

 The  mechanism in which the missing radiative corrections in the modelling
of the $Z/\gamma^* \rightarrow \tau\tau$  background could  mimic the  Higgs-decay signal was communicated 
by one of the authors of this paper (MWK) 
to the authors of  \cite{ATLAS_tau,CMS_tau} at the time of preparing these papers for publications. 
The reply was  that the other sources contributing to the uncertainties of the embedding 
procedure were dominant and the effects of the  missing QED radiative correction in the modelling
of the $Z/\gamma^* \rightarrow \tau\tau$  background are already covered by 
the corresponding systematic errors. This was claimed to be confirmed  by the Monte Carlo  studies in which 
the detector simulation results for the embedded and generated $Z/\gamma^* \rightarrow \tau\tau$
samples were compared. 

In the subsequent CMS and ATLAS technical papers, following the   $H \rightarrow \tau\tau$ 
evidence papers \cite{ATLAS_tau,CMS_tau},  
more attention has been given, by the respective groups developing the embedding procedures,
to  the importance  of  QED radiative effects. 
In the outlook section of the CMS note devoted to the embedding procedure  \cite{IEKP-KA/2015},
the importance of quantifying the impact of the final state radiation for various embedding methods 
was notified  but the corresponding studies were left for the future analyses.  
The ATLAS collaboration published very recently the paper on the embedding procedure  \cite{embedding_paper}. 
The QED radiative effects were discussed for the first time,  jointly with the resolution effects
of the reconstructed muon momenta. 
As a measure of  their importance,  the
distributions of the $\tau$  decay lepton transverse momentum and of the invariant mass of the visible
$\tau\tau$ decay products, $m^{vis}_{\tau\tau}$, for generator- and detector-seeded $\tau$ embedding
were  presented in Fig.~8a,b  of  Ref. \cite{embedding_paper}.
These  plots demonstrated that the effects of the uncorrected resolution and final-state radiation 
of the input muons are negligible in the case of reconstructed $\tau\tau$ final states, for which 
the mass resolution is dominated by the neutrinos produced in the $\tau$ decay. This conclusion is 
indeed correct\footnote{Note, that Fig.~8a  of  Ref. \cite{embedding_paper} 
is largely irrelevant to the studies of the radiative effects because hard-photon radiation affects predominantly 
the transverse momentum distribution of the sub-leading rather than the leading lepton. 
Moreover, the negative slope of the distribution of the visible $\tau\tau$ mass shown in Fig.~8b  is clearly visible even at the 5--10\% precision level of the studies presented in  \cite{embedding_paper}.
This negative slope may be considered as a footprint of the importance of the FSR effects.}, 
but only at a precision level of~5--10\%.
However, as our analysis has shown, the extrapolation  of such a general conclusion to 
the Higgs-boson searches,  where higher precision is required, cannot  be justified.

\section{The way forward}
\label{forward} 

As indicated in the previous section,   the excess of events attributed to 
the Higgs signal may  be a consequence of the missing QED final-state radiative 
correction in the modelling of  the  $Z/\gamma^* \rightarrow \tau\tau$  background. 
In  order to discriminate, beyond any doubt, between 
the three hypotheses formulated in the previous section:  FSR-only, 
Higgs-only  and FSR+Higgs, the following experimental tests may provide
additional information:
\begin{itemize} 
\item 
{\bf The  ${\mathbf m_{\tau\tau}}$ dependence test.} \\
This test boils down to selecting the event categories for which the 
experimental resolution  of the reconstructed $m_{\tau\tau}$ is the best, say
in the $10$ GeV $ \leq \sigma ^{\rm smear} _{m_{\tau\tau}} \leq 15$ GeV range,  and to performing  the 
analysis in two subsamples of events selected below and above 
$m_{\tau\tau} = 120$ GeV. If the excess of events is found to be  confined to 
the  lower mass region,  the FSR-only hypothesis is  favoured.  
If  it is found to be  concentrated in the higher mass region, the Higgs-only  hypothesis is  more likely.
\item
{\bf The embedded ${\mathbf \tau}$ momentum definition test.} \\
Here we  exploit the fact that most of the large energy photons 
radiated by the outgoing muons coming from $Z/\gamma ^*$-boson decays
are  collinear with the initial (vertex) outgoing muon direction. The identification of
FSR photons is easier thanks to  the solenoidal magnetic field in both the ATLAS
and CMS experiments. 
The FSR photon  energy clusters reconstructed in the electromagnetic calorimeter can
be differentiated 
(with a certain efficiency)  
 from the muon energy loss in the calorimeters because of the bending
 of the muons in the magnetic field.
If the energy of the identified radiative photon clusters is added to  
muon energy which is measured by its track, the size of the radiative correction to the embedding 
procedure is drastically reduced. Such a test can be 
done in particular for the VBF subsample of events, for which 
the transverse momentum spectrum of muons is softer than for 
the Boosted sample. Here the larger deflection in the magnetic
field allows for larger separation of the radiative 
photon clusters and the muon energy deposition from muon interactions
in the material in the electromagnetic calorimeter. 
By varying the association 
criteria of electromagnetic clusters with the muon, one can  experimentally control 
the level of the  radiative corrections to  the embedding procedure in the modelling
of the  $Z/\gamma^* \rightarrow \tau\tau$ background.

For higher energy muons one could add all the energy deposition
in the electromagnetic calorimeter to the energy of the muon
measured by tracking, and perform a statistical subtraction
of the energy deposition component that results from direct muon
interactions in the material of the electromagnetic calorimeter.
The statistical subtraction can utilize a spectrum of energy deposition
simulations for muons of different momenta.

\item
{\bf The bare-versus-dressed ${\mathbf \tau}$-lepton test.} \\
As shown in Fig.~\ref{fig2}, the size of the radiative corrections to the 
embedding procedure is different for these two definitions
of the outgoing $\tau$-lepton momentum.   In the majority of $\tau$-lepton decays the radiative 
photons cannot be experimentally resolved and they are included in the momentum reconstruction 
of the dressed $\tau$-lepton. However, as far as the embedding procedure
is concerned, the radiation and decays 
of $\tau$-leptons are generated by Monte Carlo 
programs in which the source of each photon is clearly defined.
The test would consist of repeating the embedding analysis twice,
for the dressed and for the bare  $\tau$-lepton definitions. If
the main source of the excess of ``Higgs-like" events are the radiative corrections
to the embedding procedure,  this excess would vary in a 
predictable way, and  the FSR-only and Higgs-only 
hypotheses could be discriminated with respect to each other. 
\item 
{\bf The selection efficiency test.} \\
The size of the radiative corrections to the embedding procedure depends very strongly 
on the relative efficiency of selecting the Higgs-like events with respect to the remaining 
background, in optimal restricted phase-space volumes. The embedding procedure,
in which the main background is determined directly from the data,  
while the Higgs signal is simulated using the Monte Carlo generated events,
is prone not only to the QED radiative corrections, but also to the  QCD radiative
corrections,  which are treated differently for data and  Monte Carlo events.
The test would consist of determining the ratio, rather than absolute values, 
of the selection efficiencies in a  restricted kinematical region from a fixed  Monte Carlo 
generator, having the same QCD approximations for the production of $Z/\gamma^*$ and Higgs bosons.
The large uncertainties of the absolute predictions
 cancel in this  ratio. 
This test is of  particular importance in view of the apparent discrepancy of the $R_{\rm FSR}^i$ 
and $R_{\rm ATLAS}^i$ values between the VBF and Boosted samples of events.
The $R_{\rm FSR}^i$ values are driven by the ratio of efficiencies, as illustrated in  Eq.~(\ref{eq:ATLASeff}),
while the $R_{\rm ATLAS}^i$ values carry, in addition, an imprint of the relative strength of the
Higgs-boson couplings to fermions and bosons. 
A single-Monte-Carlo-based measurement could be decisive for an unambiguous interpretation
of the ``VBF vs.\ Boosted" discrepancy.

\end{itemize}

\section{Conclusions}
\label{Conclusions} 

In this paper we present an analysis of the effects of  QED final-state  radiative corrections to the 
modelling of the  $Z/\gamma ^* \rightarrow \tau\tau$ process, which is the  dominant background  
in the $H \rightarrow \tau\tau$ searches. 
We focus our attention on the radiative corrections to the embedding procedure
used by ATLAS and CMS to model  the  $Z/\gamma ^* \rightarrow \tau\tau$ background in the 
extraction of the $H \rightarrow \tau\tau$ signal from experimental data. 
We make an initial estimate of the size of these corrections for  the ATLAS and CMS analyses\cite{ATLAS_tau,CMS_tau}
 and show that the missing radiative corrections in the embedding procedure  
may explain the excess of events which has been attributed to a Higgs-boson signal.
We  propose several experimental tests which can differentiate 
between an excess which originates from missing QED radiative corrections 
in the modelling of the $Z/\gamma ^* \rightarrow \tau\tau$ background
  and an excess originating from  Higgs-boson decays. 

The goal of this paper is to re-draw the attention of the CMS and ATLAS collaborations
to the importance of the effect of the  QED radiative corrections to the embedding procedure
for modelling the  $Z/\gamma ^* \rightarrow \tau\tau$ background.
Our earlier suggestions on this matter, addressed to the 
ATLAS and CMS collaborations at the time of preparation of their respective papers,  are  now quantified 
by an initial estimate  of the size of the effect. 
It is our hope that these estimates will persuade  both collaborations to 
take into account  the QED radiative corrections to their respective  embedding procedures 
used to model  the  $Z/\gamma ^* \rightarrow \tau\tau$ background in the 
extraction of the $H \rightarrow \tau\tau$ signal from current and future experimental data.

\section*{Acknowledgements}

We would like to thank T.\ Przedzi\'nski and Z.\ W\c{a}s for their help in interfacing \PHOTOS\ to \WINHAC.
We acknowledge  the help of the Academic Computer Centre CYFRONET AGH in Krakow, Poland, where 
our numerical simulations were performed with the use of the computing cluster {\sf Zeus}. We would like to 
thank Anne Pfost and Patrick Czodrowski for their critical reading of the manuscript.

\end{document}